\documentclass[prd,preprintnumbers,aps,preprint,amsmath,amssymb,superscriptaddress,floatfix,nofootinbib,unsortedaddress]{revtex4}
\pdfoutput=1

\usepackage{amssymb, bm, amsmath, graphicx, physymb}
\usepackage[colorlinks=true,linktocpage=true,linkcolor=blue,citecolor=blue]{hyperref}
\usepackage[usenames,dvipsnames]{color}
\usepackage{slashed}
\usepackage[utf8]{inputenc}
\usepackage[T1]{fontenc}
\usepackage{enumerate}
\usepackage{amsfonts}
\usepackage{mathtools}
\usepackage{latexsym}
\usepackage{hyperref}
\usepackage{epstopdf} % COMPILE EPS (AND PDF) FIGURES USING PDFLATEX!!!
\usepackage{bbm}
\usepackage{soul}
\usepackage[normalem]{ulem}
\usepackage{braket}
\usepackage{nccmath}
\usepackage{graphicx}

\usepackage[e]{esvect}

\def\cV{\mathcal{V}}

\def\cG{{\cal G}}

\def\cP{{\cal P}}
\def\cK{{\cal K}}

\def\cT{{\cal T}}

\newcommand{\secn}[1]{Section~1}
\newcommand{\appn}[1]{Appendix~1}

\long\def\comment#1{ }

\def\and{\quad\text{and}\quad}

\def\g{{\boldsymbol g}}
\def\q{{\boldsymbol q}}
\def\0{{\boldsymbol 0}}
\def\1{{\boldsymbol 1}}
\def\p{{\boldsymbol p}}
\def\l{{\boldsymbol l}}
\def\k{{\boldsymbol k}}
\def\x{{\boldsymbol x}}
\def\y{{\boldsymbol y}}

\def\r{{\boldsymbol r}}

\def\w{{\boldsymbol w}}
\def\b{{\boldsymbol b}}
\def\0{{\boldsymbol 0}}
\def\Q{{\boldsymbol Q}}

\def\rn{{\boldsymbol r}}

\def\max{{\rm max}}

\newcommand{\N}{\mathcal{N}}

\renewcommand\a{\alpha}
\renewcommand\b{\beta}
\renewcommand\d{\delta}
\renewcommand\l{\lambda}
\renewcommand\r{\rho}

\renewcommand\t{\tau}

\renewcommand\o{\omega}
\newcommand\e{\epsilon}
\newcommand\m{\mu}
\newcommand\n{\nu}

\renewcommand\L{\Lambda}

\renewcommand\O{\Omega}

\newcommand\G{\Gamma}

\newcommand{\re}{{\rm{Re}}}

\newcommand{\tvec}{\boldsymbol}
\renewcommand{\vec}{\boldsymbol}

\renewcommand{\part}{{\rm part}}
\newcommand{\be}{\begin{equation}}
\newcommand{\ee}{\end{equation}}
\newcommand{\bes}{\begin{subequations}}
\newcommand{\ees}{\end{subequations}}
\newcommand{\bea}{\begin{eqnarray}}
\newcommand{\eea}{\end{eqnarray}}

\newcommand{\pa}{\partial}

\newcommand{\nn}{\nonumber \\}
\newcommand{\na}{\nabla}

\begin{document}

\title{Medium induced gluon spectrum in dense inhomogeneous matter}

%--------------------------------------------------------------------------
\author{Jo\~{a}o Barata}
\email[Email: ]{jlourenco@bnl.gov}
\affiliation{Physics Department, Brookhaven National Laboratory, Upton, NY 11973, USA}
%--------------------------------------------------------------------------
\author{Xo{\'{a}}n Mayo L\'{o}pez}
\email[Email: ]{xoan.mayo.lopez@usc.es}
\affiliation{Instituto Galego de F{\'{i}}sica de Altas Enerx{\'{i}}as,  Universidade de Santiago de Compostela, Santiago de Compostela 15782,  Galicia, Spain}
%--------------------------------------------------------------------------
\author{Andrey V. Sadofyev}
\email[Email: ]{andrey.sadofyev@usc.es}
\affiliation{Instituto Galego de F{\'{i}}sica de Altas Enerx{\'{i}}as,  Universidade de Santiago de Compostela, Santiago de Compostela 15782, Galicia, Spain}
%--------------------------------------------------------------------------
\author{Carlos A. Salgado}
\email[Email: ]{carlos.salgado@usc.es}
\affiliation{Instituto Galego de F{\'{i}}sica de Altas Enerx{\'{i}}as,  Universidade de Santiago de Compostela, Santiago de Compostela 15782,  Galicia, Spain}

%--------------------------------------------------------------------------

\begin{abstract}
 We calculate the spectrum of gluons sourced by the branching of an energetic quark in the presence of an inhomogeneous QCD medium, focusing on the soft radiation limit. We take into account multiple soft interactions between the partons and matter, treating the transverse variations of its parameters within a gradient expansion. Thus, we derive the general form of the medium induced spectrum up to the first order in gradients, and consider its simplifying limits. In particular, we show that to the leading order in matter gradients and using the harmonic approximation for the scattering potential, the full gluon spectrum can be written in a compact closed form suitable for numerical evaluation. The final gluon transverse momentum tends to align along the anisotropy direction, resulting in a non-trivial azimuthal pattern in the jet substructure. 
\end{abstract}

\maketitle

\section{Introduction}

Over the last decades, the experiments on high-energy heavy-ion collisions (HIC) at RHIC and the LHC allowed to explore QCD at high energies and densities, for a review see e.g. \cite{Busza:2018rrf, Apolinario:2022vzg}. In these experiments, the nuclear matter is produced far from equilibrium, and undergoes a multiphase evolution. After the initial non-equilibirum dynamics, the matter thermalizes into a nearly ideal liquid, the quark-gluon plasma (QGP), which continues expanding and cooling. When the energy density is low enough, the matter turns into a hadron gas, which is eventually observed by the detectors. Following the initial observation of the QGP formation, the main community efforts have been concentrated on extracting the details of the matter evolution in HIC. 

One of the evidences of the collective matter formation in HIC is the suppression of energetic partons due to the in-medium energy loss~\cite{STAR:2002ggv,ALICE:2010yje,PHENIX:2001hpc}, known as jet quenching~\cite{Bjorken:1982tu}, for recent reviews see \cite{Mehtar-Tani:2013pia,Qin:2015srf,Blaizot:2015lma}. Moreover, the cascades of secondary particles produced by the branching of such energetic partons, forming jets, are also modified by the matter, providing a differential tool to probe the medium at different length and energy scales. Using jets for such imaging of the nuclear matter and its evolution in HIC and other experiments is often referred to as jet tomography, see e.g. \cite{Vitev:2002pf, Wang:2002ri, Xu:2014ica, Djordjevic:2016vfo, Apolinario:2017sob, Arratia:2020nxw, He:2020iow, Apolinario:2020uvt, Sadofyev:2021ohn, Du:2021pqa, Hauksson:2021okc, Antiporda:2021hpk,Sadofyev:2022hhw,Andres:2022ndd,Fu:2022idl,Barata:2022wim,Hauksson:2023tze,Boguslavski:2023alu} and references therein.

Considering the jet-medium interactions in perturbative QCD, one usually has to describe the underlying hot matter in terms of a background stochastic color field, see e.g.~\cite{Gyulassy:1993hr, Zakharov:1996fv, Baier:1996kr,  Wiedemann:2000ez, Wiedemann:2000za, Gyulassy:2000er, Gyulassy:2002yv, Arnold:2002ja}. Moreover, to make the calculations tractable, most works further rely on multiple simplifying assumptions. For instance, the medium is commonly considered to be transversely homogeneous with a finite longitudinal extension,\footnote{The longitudinal and transverse directions are defined with respect to the initial momentum of the leading parton momentum.} while the calculations are preformed in the large energy limit, known as the eikonal approximation. Under these assumptions, the problem allows for a semi-analytic treatment, but the results cannot be applied to resolve the details of the medium evolution, and jets appear to be decoupled from the anisotropic matter expansion \cite{Sadofyev:2021ohn} and, moreover, from the large anisotropies of the initial out-of-equilibrium phase of the matter produced in HIC, see e.g. \cite{Sadofyev:2021ohn, Hauksson:2021okc, Carrington:2021dvw, Ipp:2020mjc}. Only recently, the theory of jet-matter interactions has been extended to the case of inhomogeneous nuclear matter~\cite{Fu:2022idl,Sadofyev:2021ohn,Barata:2022krd,Barata:2022utc} with the transverse matter anisotropies treated within a gradient expansion.\footnote{See also \cite{Lekaveckas:2013lha, Rajagopal:2015roa, Sadofyev:2015hxa, Li:2016bbh, Reiten:2019fta, Arefeva:2020jvo} for applications of the same gradient expansion approach in holographic models of probe-matter interactions.} 

So far, the approach developed in \cite{Sadofyev:2021ohn,Barata:2022krd} has been used to describe single parton evolution, the so-called jet broadening, in dilute or dense media at the leading order in transverse gradients, but the case of in-medium branching in inhomogeneous matter has not been considered. In this paper, we continue developing the formalism by deriving the medium induced soft gluon spectrum up to the first order in transverse gradients but to all orders in opacity. Our main result is the double differential spectrum, which can be written as
\begin{align}
\label{spectrumintro}
&\omega\frac{dI}{d\omega d^2\k}= \omega\frac{dI_0}{d\omega d^2\k}+(\hat{\g}\cdot \k) \,\omega\frac{dI_1}{d\omega d^2\k} +\mathcal{O}(\hat{\g}^2)\, ,
\end{align}
where $dI_0$ denotes the gluon spectrum in homogeneous matter, $dI_1$ gives the functional dependence of the leading order gradient contribution on $\k^2$, resulting in a non-trivial azimuthal dependence of the overall spectrum, and $\o$ and $\k$ are the energy and momentum of the emitted gluon. Notice that in this work we have derived \eqref{spectrumintro} in the approximation of static matter (no flow), although the generalization for flowing medium is straightforward to obtain. As in \cite{Barata:2022krd}, we use the Gyulassy-Wang (GW) model to make some of the expressions explicit, and the primary matter parameters are the number density of the scattering centers $\rho$ and Debye mass $\mu$. However, our results are general and can be directly extended to other models for the source potential. The corresponding transverse gradients are encapsulated in a two-dimensional vector operator $\hat \g \equiv \left(\vec{\na}\rho \frac{\delta }{\delta \rho} + \vec{\na }\mu^2 \frac{\delta }{\delta \mu^2}  \right)$. Below, we will show that the gradient terms in the spectrum result in a deflection of the emitted gluons along $\hat \g$. 

This paper is organized as follows: In section~\ref{sec:medium_spectrum} we derive the generic form of the medium induced gluon spectrum up to first order in transverse gradients for media with finite longitudinal extension. In section~\ref{sec:medium_spectrum_harmonic}, we further simplify the generic form of the spectrum, utilizing the harmonic approximation limit for the in-medium scattering potential. In section~\ref{sec:gradients}, we obtain the gradient corrections to the medium induced gluon spectrum in the considered limit. We discuss the properties of the spectrum in section~\ref{sec:results}, as well as provide some numerical results, illustrating its behavior. Finally, our findings are summarized in section~\ref{sec:conclusion}, where we also discuss potential extensions of the presented results.

\section{Medium induced spectrum in the soft gluon limit}\label{sec:medium_spectrum}

\subsection{Resummation at the amplitude level}

It is instructive to briefly repeat the derivation of the amplitude describing the propagation of a quark with initial energy $E$ and transverse momentum $\p_{1}$ in a medium, followed by an emission of a gluon, which is measured to have energy $\omega$ and transverse momentum $\k$ in the final state. Following \cite{Sadofyev:2021ohn}, we will focus on the transverse gradients of the source density and Debye mass, assuming the matter to be static. We also assume the in-matter sources to be static by themselves, ignoring such effects as the medium response and collisional energy loss, which would significantly complicate our consideration, and require futher generalization of the formalism developed in \cite{Sadofyev:2021ohn,Barata:2022krd}. Thus, the medium induced color field can be chosen as 
\bea
\label{eq:A_mu}
g A^{a\l}(q) = (2\pi)\,g^{\l0}\,v^a(q)\,\d\left(q^0\right)\,,
\eea
and the jet-medium interactions are controlled by
\bea
v^a(q)=\int_{x}\, e^{-i\left(\tvec{q}\cdot\tvec{x}+q_z z\right)}\hat\rho^a(\tvec{x}, z)\,v(q,\vec{x},z)\,,
\eea
where $\hat \rho^a$ is the source color density, $v(q,\vec{x},z)$ is a single source potential, which depends on coordinates through the local medium properties. We have also introduced shorthand notations for integrals running over the full three-dimensional space as $\int_x\equiv d^3 x$ and $\int_k \equiv \frac{d^3 k}{(2\pi)^3}$, and over the transverse space as  $\int_{\x} \equiv \int d^2\x$ and $\int_{\k} \equiv \int \frac{d^2 \k }{(2\pi)^2}$. 

The single source potential is expected to be exponentially screened in coordinate space, and we will generally refer to the screening scale as the Debye mass. While the medium induced soft gluon spectrum can be derived for the general potential, and we will do so here, it is instructive to consider an explicit form of $v(q,\tvec{x},z)$ as well. For this purpose, we will follow \cite{Barata:2022krd}, and refer to the GW model \cite{Gyulassy:1993hr}, corresponding to
\be
v(q,\vec{x},z) = \frac{g^2}{q^2-\mu^2(\vec{x},z)+i\epsilon}\,.
\label{eq:potential}
\ee
Here, we assume that $\mu(\vec{x},z)$ varies slowly in the transverse direction over distances of order $\frac{1}{\m({\bf 0},z)}$. Focusing on the limit, when the characteristic distance between the sources is larger than $\frac{1}{\m({\bf 0},z)}$, we ignore the local modifications in general $v(q,\vec{x},z)$ for the given source, taking into account only the changes in the Debye mass between different sources.   

\begin{figure}[h]
    \centering
    \includegraphics[width=0.8\textwidth]{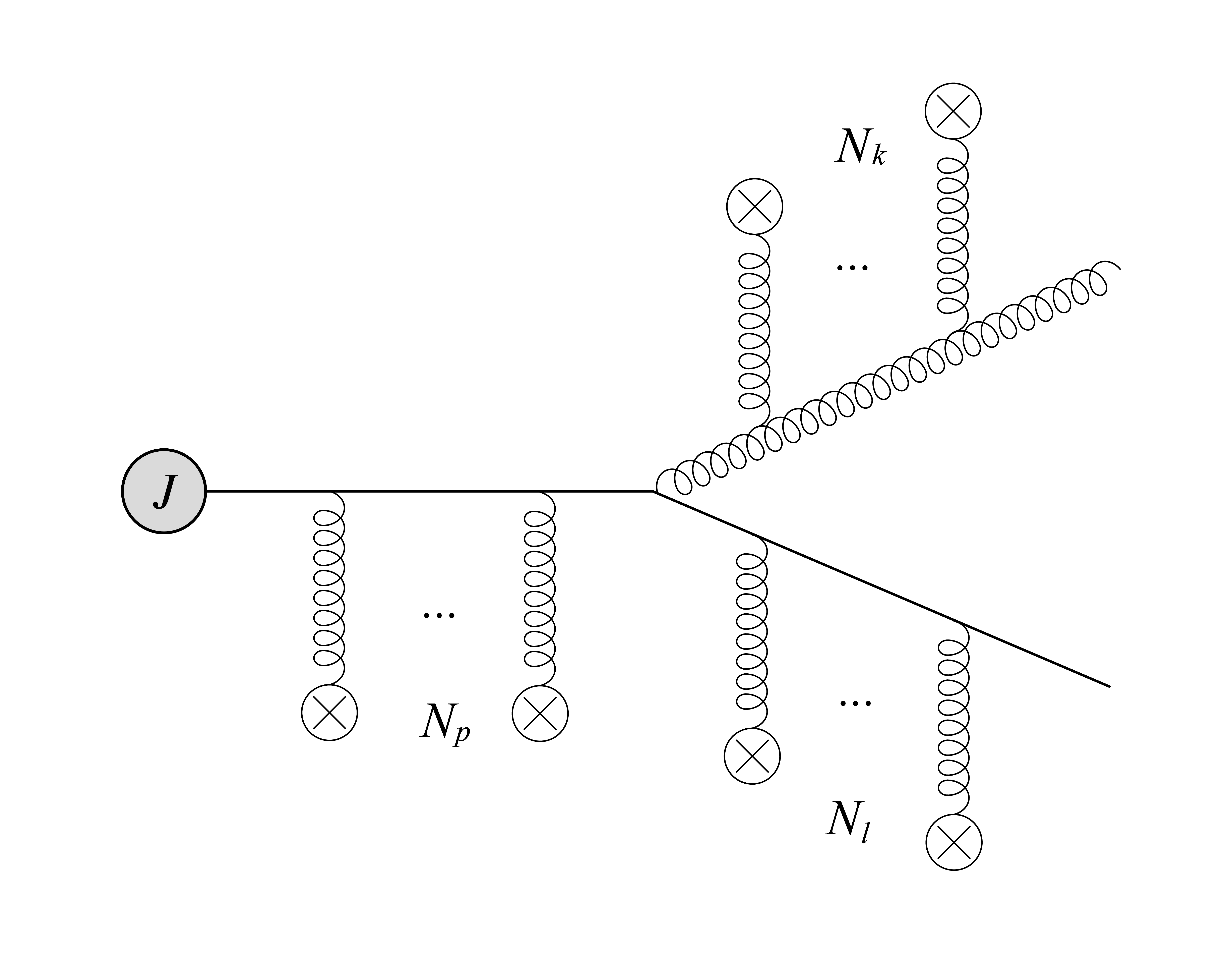}
    \caption{Feynman diagram of the in-medium gluon emission from a quark with each particle interacting multiple times with the background field.}
    \label{fig:diagram}
\end{figure}

Using this model for the medium, we can write the amplitude, $iR_{N_p N_l N_k}$, depicted in Fig.~\ref{fig:diagram}, with $N_p$ insertions in the incoming quark line, $N_l$ insertions in the outgoing quark line, and $N_k$ insertions in the gluon line, as
\begin{align}
\label{Rfull}
    i\mathcal{R}_{N_p N_l N_k} &=\prod_{n=1}^{N_p} \left[(-1)\int_{p_n} t^{a_n}_{proj} v^{a_n}(p_{n+1}-p_n) 
    \frac{2E}{p_n^2+i\epsilon}\right] \notag\\
     & \times \int \frac{d^4p_s}{(2\pi)^4}\,  (-1)\,gt^{b_1}_{proj}\frac{(p_s+l_{1})_{\mu_1}}{p_s^2+i\epsilon}\,(2\pi)^4 \delta^{(4)}(p_s-l_{1}-k_{1}) J(p_1)\notag\\
    &\times \prod_{r=1}^{N_k} \left[\left(-\frac{i}{g}\right)\int_{k_r} \frac{N^{\mu_r \nu_r}(k_r)}{k_r^2+i\epsilon} 
    \Gamma^{b_r b_{r+1} c_r}_{\nu_r \mu_{r+1} 0}(k_r,-k_{r+1}) \,v^{c_r}(k_{r+1}-k_r)\right]\epsilon^{*\mu_{N_k+1}}(k)\notag\\
    & \times  \prod_{m=1}^{N_l} \left[(-1)\int_{l_m} t^{d_m}_{proj}v^{d_m}(l_{m+1}-l_m)
    \frac{2(1-x)E}{l_m^2+i\epsilon}\right],
\end{align}
where $\G^{abc}_{\a\b\gamma}$ is the three-gluon vertex, the zeroth components of the four-momenta have already been fixed with the delta functions coming from the background field insertions, and we have introduced additional momentum labels $p_{N_p+1}=p_s$, $l_{N_l+1}=l$, and $k_{N_k+1}=k$. Notice that the integrations in \eqref{Rfull} should be understood as acting from the left on the whole expression, and are distributed now for structural simplicity.

This expression can be simplified if we use the particular kinematics of the process. Indeed, let us take a closer look at the first product in \eqref{Rfull}. Switching to the Fourier transformed in-medium potentials, and re-ordering the Fourier exponentials within the product, we can write it as
\begin{align}
&\prod_{n=1}^{N_p} \Bigg[(-1)\int_{p_n,x_n} t^{a_n}_{proj} \hat{\r}^{a_n}(\vec{x}_n,z_n)\, v(p_{n+1}-p_n,\vec{x}_n,z_n)\notag\\
&\hspace{5cm}\times\frac{2E}{p_n^2+i\epsilon}e^{-i\vec{x}_n\cdot(\vec{p}_{n+1}-\vec{p}_n)}e^{ip_{n,z}(z_n-z_{n-1})}\Bigg]e^{-ip_{s,z}z_{N_p}}J(p_1)\,,
\end{align}
where we have introduced $z_0=0$. One can further perform $p_{n,z}$ integrations by residues, collecting the corresponding poles. We also assume that $\m \Delta z\gg1$, where $\Delta z$ is the characteristic distance between the color sources (described by $\r^a$), and that the single source potential has no other poles apart from the screening ones. Thus, we neglect the poles of the in-medium potentials, which are exponentially suppressed in the considered limit. Assuming that the matter is extended in the positive $z$-direction, we find that only the poles of the scalar propagators with $p_{n,z}>0$ contribute. Finally, we take the limit of small gluon energy fraction $x=\frac{\o}{E}\ll1$, and neglect all the subeikonal terms unless they are enhanced by the medium length and $\frac{1}{x}$ simultaneously. Under these approximations, we find
\begin{align}
\label{intNp}
    \prod_{n=1}^{N_p} \left[i\int_{\mathbf{p}_n,x_n} t^{a_n}_{proj} \tilde{v}^{a_n}(\vec{x}_n,z_n)\,\theta_{n,n-1}\,e^{-i\vec{x}_n\cdot(\vec{p}_{n+1}-\vec{p}_n)}\right]e^{-i(p_{s,z}-E)z_{N_p}}J(p_1)\,,
\end{align}
where $\tilde{v}^a(\vec{x},z)=\int_{\q,\y}e^{i\q\cdot (\x-\y)}\hat{\r}^a(\vec{y},z)\, v(q,\vec{y},z)\big|_{q_0=q_z=0}$. Fourier transforming the momentum conservation in the emission vertex and substituting \eqref{intNp}, we can integrate over $p_{s,z}$ in \eqref{Rfull}. Thus, we set $p_{s,z}\simeq E$ and impose an upper limit on variations of $z_{N_p}$. Now one may sum \eqref{intNp} over all possible diagrams, treating the case of $N_p=0$ explicitly,
\begin{align}
    &J(p_s)+\sum_{N_p=1}^\infty\prod_{n=1}^{N_p} \left[i\int_{\mathbf{p}_n,x_n} t^{a_n}_{proj} \tilde{v}^{a_n}(\vec{x}_n,z_n)\,\theta_{n,n-1}\,e^{-i\vec{x}_n\cdot(\vec{p}_{n+1}-\vec{p}_n)}\right]\theta_{sN_p}J(p_1)\notag\\
    &\hspace{3cm}=\int_{\mathbf{p}_1,\vec{x}_1}e^{-i\vec{x}_1\cdot(\vec{p}_s-\vec{p}_1)}\cP\exp\left\{i\int_0^{z_s}\,d\t\,t^a_{proj}\tilde{v}^a(\vec{x}_1,\t)\right\}J(p_1)\, ,
\end{align}
where the Wilson line of the scattering potential in the fundamental representation can be identified
\begin{equation}
    \mathcal{W}(\vec{x}_1;z_s,0) = \cP\exp\left\{i\int_0^{z_s}\,d\t\,t^a_{proj}\tilde{v}^a(\vec{x}_1,\t)\right\}\,.
\end{equation}
Similarly, looking at the other quark leg in the fourth line of \eqref{Rfull}, we readily find
\begin{align}
   &(p_s+l)_{\m_1}+\sum_{N_l=1}^\infty\prod_{m=1}^{N_l} \left[(-1)\int_{l_m} t^{d_m}_{proj} v^{d_m}(l_{m+1}-l_m)
    \frac{2(1-x)E}{l_m^2+i\epsilon}\right]e^{-i(l_{1,z}-(1-x)E)z_s}(p_s+l_{1})_{\m_1}
   \notag\\
   &\hspace{2cm}=
   \int_{\vec{l}_1,\vec{x}_1} e^{-i\vec{x}_1\cdot(\vec{l}-\vec{l}_1)}\cP\exp\left\{i\int_{z_s}^\infty\,d\t\,t^d_{proj}\tilde{v}^d(\vec{x}_1,\t)\right\}(p_s+l_{1})_{\m_1}\, .
\end{align}
where\footnote{It should not be confused with $z_0$ in the initial quark line at $n=0$.} $z_0=z_s$, the exponential $e^{-il_{1,z}z_s}$ is coming from the emission vertex, and the final momentum of the quark $l_{N_l + 1}=l$ is on shell.

Now, we have to consider the third type of insertions, attached to the emitted gluon line. Its structure is more involved, and it is instructive to make several simplifying observations upfront. First, one has to specify the gauge choice, fixing the form of $\e^{*\m}$ and $N^{\m\n}$. The background field \eqref{eq:A_mu}
induced by the matter sources is derived in the Lorentz gauge. However, working in this gauge would considerably complicate the consideration, since the polarization vector would involve auxiliary components. To remove the residual gauge freedom, one should notice that the particular form of \eqref{eq:A_mu} is compatible with an additional gauge condition $A^z=0$. In this gauge, combining the two constraints together, we can write the polarization vector in terms of physical components 
\be
\e^*_\m(k)=\left\{\frac{\vec{\e}\cdot\vec{k}}{\o}, \vec\e, 0\right\}\,,
\ee
while the numerator of the gluon propagator reads
\be
N^{\m\n}(k)=g^{\m\n}-\frac{k^\m n^\n + k^\n n^\m}{(k\cdot n)}-\frac{k^\m k^\n}{(k\cdot n)^2}\,,
\ee
where we have introduced the 4-vector $n_\mu = \{0,0,0,1\}$ entering through the gauge condition $n\cdot A=0$. Notice that we give the on-shell form of $N^{\m\n}$, since $k_{n,z}$ will be set on-shell by the corresponding integrations.\footnote{The propagator numerators lead to no new poles, leaving the $k_z$-integrations unaffected.} Thus, one readily finds
\begin{align}
\label{NGamma}
&N^{\mu_r \nu_r}(k_r)\Gamma^{b_r b_{r+1} c_r}_{\nu_r \mu_{r+1} 0}(k_r,-k_{r+1})\epsilon^{*\mu_{r+1}}(k_{r+1}) \simeq 2\o gf^{b_r b_{r+1} c_r}\e^{*\m_r}(k_r)\,,
\end{align}
where we have neglected the subeikonal terms. 

Noticing that $\left(T^c\right)_{ab}=-if^{abc}$ and using \eqref{NGamma}, we can re-express the second line in \eqref{Rfull}, adding the contribution $N_k=0$ explicitly, as
\begin{align}
\label{Rthegluonleg}   
& e^{i\frac{\vec{k}^2}{2\o}z_s} \, \d^{b_{N_k +1} b_1} \,\epsilon^{*\mu_1}(k)+\sum_{N_k=1}^\infty\prod_{r=1}^{N_k} \left[ i\int_{\k_r,x_r} \left(T^{c_r}\right)_{b_{r+1} b_r} \tilde{v}^{c_r}(\vec{x}_r,z_r)\theta_{r,r-1}e^{-i\vec{x}_r\cdot(\vec{k}_{r+1}-\vec{k}_r)} e^{iQ_r(z_r-z_{r-1})}\right]\,\notag\\
& \hspace{2.25cm} \times e^{-ik_{z}z_{N_k}+i\o z_s}\,\epsilon^{*\mu_1}(k_1) = \lim_{z_f\to\infty}\int_{\k_1} e^{i\frac{\vec{k}^2}{2\o}z_f}\,\mathcal{G}^{b_{N_k +1} b_1}\left(\vec{k},z_f; \vec{k}_{1}, z_s\right)\,\epsilon^{*\mu_1}(k_{1})\,,
\end{align}
where we keep the leading subeikonal contributions to the poles of the scalar propagators $Q_r=\o-\frac{\k^2_{r}}{2\o}$, resulting in the so-called Landau-Pomeranchuk-Migdal (LPM) phases, which are enhanced in the soft gluon limit by $\frac{1}{x}$. Here, we have introduced a single-particle propagator, which has the following coordinate space representation
\begin{align}
    \mathcal{G}\left(\vec{x}_f,z_f; \x_{in}, z_{in}\right)&=\int^{\vec{x}_f}_{\x_{in}}\mathcal{D}\vec{r}\,\exp\left\{\frac{i\o}{2}\int_{z_{in}}^{z_{f}}d\t\,\dot{\vec{r}}^2\right\}\notag\\
    &\hspace{1.5cm}\times\cP\exp\left\{i\int^{z_f}_{z_{in}}d\t\,T^c\tilde{v}^c(\vec{r}(\t),\t)\right\}\,,
\end{align}
where $z_f$ has been introduced in \eqref{Rthegluonleg} to simplify the definition of $\cG$.

Combining these three contributions and introducing $\vec{x}_{in}\equiv\vec{x}_1$, we find that in the soft gluon limit, the resummed amplitude can be written as
\begin{align}
\label{Rsimple}
    i\mathcal{R} &
    \simeq-\frac{g}{\o}\lim_{z_f\to\infty}\int_0^\infty dz_s\,\int_{\x_{in}} e^{-i\x_{in}\cdot\vec{l}}J(\x_{in})\notag\\ &\hspace{1cm} \times  \mathcal{W}(\x_{in}; \infty, z_s)\,t^a_{proj}\,\mathcal{W}(\x_{in}; z_s, 0)\,\,e^{i\frac{\vec{k}^2}{2\o}z_f}\,\left[\vec{\e}\cdot\vec{\na}_{\vec{x}_{in}}\mathcal{G}^{ba}\left(\vec{k},z_f; \x_{in}, z_s\right)\right]\,,
\end{align}
where the color indices have been renamed for simplicity.

\subsection{Medium average}

With the explicit form of the resummed amplitude, we can now turn to the medium averaging, required to describe the medium induced branching. The final state gluon distribution $\N$ is defined as
\begin{align}
2(2\pi)^3 \omega E \frac{d\N}{d\omega dE d^2\k}\equiv   \frac{1}{4\pi} \frac{1}{N_c}\sum\int \frac{d^2 \boldsymbol{l}}{(2\pi)^2}\,\langle|\mathcal{R}|^2\rangle\,, 
\end{align}
where we have averaged over the stochastic background field. Thus, upon squaring \eqref{Rsimple}, summing over final state quantum numbers, and averaging over the initial ones, we find that
\begin{align}\label{eq:M_1}
    2(2\pi)^3 \omega E \frac{d\N}{d\omega dE d^2\k} &=\lim_{z_f\to\infty}\frac{\alpha_s}{N_c\, \omega^2}\int_0^\infty dz\, \int_0^\infty d\bar{z}\,\int_{\x_{in}} |J(\x_{in})|^2\notag\\ 
    &\hspace{-2.5cm} \times  \Bigg\langle\text{Tr}\Bigg[\mathcal{W}(\x_{in}; \infty, z)\,t^a_{proj}\,\mathcal{W}(\x_{in}; z, 0)\,\left[\na_{\a, \x_{in}}\mathcal{G}^{ba}\left(\vec{k}_f,z_f; \x_{in}, z\right)\right]\notag\\ 
    &\hspace{-1.5cm} \times  \Big(\mathcal{W}(\x_{in}; \infty, \bar{z})\,t^{\bar{a}}_{proj}\,\mathcal{W}(\x_{in}; \bar{z}, 0)\,\left[\na_{\a, \x_{in}}\mathcal{G}^{b\bar{a}}\left(\vec{k},z_f; \x_{in}, \bar{z}\right)\right]\Big)^\dag\Bigg]\Bigg\rangle\, .
\end{align}
where the subscript $\alpha$ in the derivatives runs over the two components of the transverse vector $\x_{in}$.

This expression can further be simplified if one notices that the fundamental Wilson lines can be combined to form the Wilson line in the adjoint representation, see e.g. \cite{Casalderrey-Solana:2007knd, Mehtar-Tani:2012mfa,Blaizot:2012fh}. Indeed, if $\bar{z}>z$, then  
\begin{align}
    &\text{Tr}\Big[\mathcal{W}(\vec{x}_{in}; \infty, z)\,t^a_{proj}\,\mathcal{W}(\vec{x}_{in}; z, 0)\mathcal{W}^\dag(\vec{x}_{in}; \bar{z}, 0)\,t^{\bar{a}}_{proj}\, \mathcal{W}^\dag(\vec{x}_{in}; \infty, \bar{z})\Big]\notag\\
    &\hspace{1cm}=\text{Tr}\Big[\mathcal{W}(\vec{x}_{in}; \bar{z}, z)\,t^a_{proj}\,\mathcal{W}^\dag(\vec{x}_{in}; \bar{z}, z)\,t^{\bar{a}}_{proj}\Big]\notag\\
    &\hspace{2cm}=\frac{1}{2}\mathcal{W}^{\dag a\bar{a}}_A(\vec{x}_{in};\bar{z},z)\,,
\end{align}
which in terms of the adjoint generators reads
\begin{equation}
    \mathcal{W}_A^{\dagger a\bar{a}}(\x_{in};\bar{z},z) = \cP \exp\left\{-i\int_z^{\bar{z}} d\tau \left(T^c\right)^{a\bar{a}} \tilde{v}^c(\x_{in},\tau)\right\}
\end{equation}
and we can rewrite \eqref{eq:M_1} as
\begin{align}\label{eq:M_2}
    2(2\pi)^3\omega E \frac{d\N}{d\omega dE d^2\k} &=\lim_{z_f\to\infty}\frac{\alpha_s}{N_c\,\omega^2}\,\re \int_0^\infty d\bar{z}\int_0^{\bar{z}} dz\,\int_{\x_{in}} \,|J(\x_{in})|^2\notag\\ 
    &\hspace{-2cm} \times  \left\langle\left[\na_{\a,\x_{in}}\mathcal{G}^{ba}\left(\vec{k},z_f; \x_{in}, z\right)\right]\mathcal{W}_A^{\dag\,a\bar{a}}(\x_{in}; \bar{z},z)\left[\na_{\a, \x_{in}}\mathcal{G}^{\dag\,\bar{a}b}\left(\vec{k},z_f; \x_{in}, \bar{z}\right)\right]\right\rangle \,,
\end{align}
where the contributions from the two regions with $\bar{z}>z$ and $z>\bar{z}$ combine into the real part of the expression above.

We further assume that the color source densities have Gaussian statistics, enforcing the color neutrality condition, see e.g. \cite{Andres:2022ndd}. Then, the only non-trivial average is given by
\be
\label{eq:bgfield}
\left\langle\hat\rho^a(\vec{x}, z)\hat\rho^b(\bar{\vec{x}}, \bar{z})\right\rangle = \frac{1}{2C_{\bar R}}\delta^{ab}\,\delta^{(2)}(\vec{x}-\bar{\vec{x}})\,\delta(z-\bar{z})\,\rho(\vec{x}, z)\,,
\ee
or, equivalently,
\be
\left\langle\hat\rho^a(\vec{q},z)\hat\rho^b(\bar{\vec{q}},\bar{z})\right\rangle 
= \frac{1}{2C_{\bar R}}\delta^{ab}\delta(z-\bar{z})
\int_\x \, e^{-i(\vec{q}+\bar{\vec{q}})\cdot\vec{x}}\,\rho(\vec{x},z)\,,
\label{eq:FTaverage}
\ee
where $\rho(\x,z)$ denotes the number density of the scattering centers in the medium, the sources are assumed to be in the same representation $R$, and $C_{\bar{R}}$ is the quadratic Casimir in the representation opposite to $R$. In coordinate space, $\hat\rho^a(\vec{x}, z)$ is real, and its Fourier transform satisfies  $\hat\rho^{a\dagger}(\vec{q},z) = \hat\rho^a(-\vec{q},z)$. Thus,
\begin{align}
\label{vvint}
\langle \tilde{v}^a(\vec{x},z)\tilde{v}^{\dagger b}(\bar{\vec{x}},\bar{z})\rangle
&\simeq\frac{\d^{ab}}{2C_{\bar R}}\,g^4\,\d(z-\bar z)\int_{\q,\Q}
e^{i\vec{q}\cdot(\vec{x}-\bar{\vec{x}})}\r(z)
\,(2\pi)^2 \d^{(2)}(\Q) \notag\\
&\hspace{-2.5cm}\times\left[1+  \hat{\vec g} \cdot\left(\frac{\vec{x}+\bar{\vec{x}}}{2} + \frac{\partial}{\partial\Q}\right)\right]
v\left(\q+\frac{1}{2}\Q,0,z\right) v^\dagger\left(\q-\frac{1}{2}\Q,0,z\right)\bigg|_{q_0=q_z=Q_0=Q_z=0} \,,
\end{align}
where the leading gradient corrections are accounted by $\hat{\vec{g}}$, see e.g. \cite{Barata:2022krd}. Taking the GW model as an illustrative example, this average reads 
\begin{align}
\label{vvint}
\langle \tilde{v}^a(\vec{x},z)\tilde{v}^{\dagger b}(\bar{\vec{x}},\bar{z})\rangle
\simeq\frac{\d^{ab}}{2C_{\bar R}}\,g^4\,\d(z-\bar z)\left(1+\frac{\vec{x}+\bar{\vec{x}}}{2}\cdot\hat{\vec g}\right)\int_{\q}  \frac{\r(z)\,e^{i\vec{q}\cdot(\vec{x}-\bar{\vec{x}})}}{\left(\q^2+\m^2(z)\right)^2} \,.
\end{align}

Even in the presence of transverse gradients, the average \eqref{vvint} is still local in $z$. Consequently, an average of a product of single-particle propagators with no common support reduces to a product of averages. That allows us to write \eqref{eq:M_2} as
\begin{align}
\label{RRavr}
    2(2\pi)^3 \omega E \frac{d\N}{d\omega dE d^2\k} &=\lim_{z_f\to\infty}\frac{\alpha_s}{N_c\, \omega^2}\re \int_0^\infty d\bar{z}\int_0^{\bar{z}} dz\,\int_{\x_{in}, \y} \,|J(\x_{in})|^2\,\Bigg[\left(\vec{\na}_\x\cdot\vec{\na}_{\bar \x}\right)\notag\\ 
    &\hspace{-3cm} \times  \Big\langle\mathcal{G}^{bc}\left(\vec{k},z_f; \vec{y}, \bar{z}\right)\mathcal{G}^{\dag\,\bar{a}b}\left(\vec{k},z_f; \bar{\vec{x}}, \bar{z}\right)\Big\rangle\Big\langle\mathcal{G}^{ca}\left(\vec{y},\bar{z}; \vec{x}, z\right)\mathcal{W}_A^{\dag\,a\bar{a}}(\x_{in};  \bar{z},z)\Big\rangle\Bigg]\Bigg|_{\vec{x}=\bar{\vec{x}}=\x_{in}} \,,
\end{align}
where we have used a mixed coordinate representation. 

Due to the color triviality of the medium averages, one can simplify the color structure in \eqref{RRavr}, writing the full process in terms of an emission kernel, $\cK$, with support in the interval $(z,\bar z)$, and a broadening kernel, $S_2$, which describes the evolution of the radiated gluon after being produced. These kernels are defined to be 
\begin{align}
S_2\left(\vec{k},\vec{k},z_f; \vec{y}, \bar{\vec{x}}, \bar{z}\right)&=\frac{1}{N_c^2-1}\Big\langle\mathcal{G}^{bc}\left(\vec{k},z_f; \vec{y}, \bar{z}\right)\mathcal{G}^{\dag\,cb}\left(\vec{k},z_f; \bar{\vec{x}}, \bar{z}\right)\Big\rangle \, ,\notag\\
\mathcal{K}\left(\vec{y},\vec{x}_{in},\bar{z}; \vec{x},\vec{x}_{in}, z\right)&=\frac{1}{N_c^2-1}\Big\langle\mathcal{G}^{ba}\left(\vec{y},\bar{z}; \vec{x}, z\right)\mathcal{W}_A^{\dag\,ab}(\vec{x}_{in};  \bar{z},z)\Big\rangle\, ,
\end{align}
such that the full distribution can be compactly written as
\begin{align}
    2(2\pi)^3 \omega E \frac{d\N}{d\omega dE d^2\k} &=\frac{2\alpha_sC_F}{\omega^2}  \re \int_0^\infty d\bar{z}\int_0^{\bar{z}} dz\,\int_{\x_{in},\y}\,|J(\vec{x}_{in})|^2\,\Bigg[\vec{\na}_\x\cdot\vec{\na}_{\bar \x}\notag\\ 
    &\hspace{-2cm} \times  S_2\left(\vec{k},\vec{k},\infty; \vec{y}, \bar{\vec{x}}, \bar{z}\right) \mathcal{K}\left(\vec{y},\vec{x}_{in},\bar{z}; \vec{x},\vec{x}_{in}, z\right)\Bigg]\Bigg|_{\vec{x}=\bar{\vec{x}}=\vec{x}_{in}} \, .
\end{align}
Thus, the final soft gluon distribution factorizes into the two correlators even in the presence of transverse gradients, while the effects of the matter anisotropy enter each one of them independently.

Now, we can utilize the explicit form of the two-point correlator \eqref{vvint}, taking into account the gradient corrections. Here, we will focus on the case of longitudinally uniform matter, keeping the algebra more compact, while the results can be straightforwardly generalized. The corresponding broadening two-point function has already been derived in \cite{Barata:2022krd} up to the first order in transverse gradients, and its in-medium part reads
\begin{align}
\label{S2}
S_2\left(\vec{k},\vec{k},L; \vec{y}, \bar{\vec{x}},\bar{z}\right)&\simeq e^{-i\tvec{k}\cdot \vec{u}}e^{-\mathcal{V}(\vec{u})(L-\bar{z})}\Bigg[1+\frac{i(L-\bar{z})^3}{3\o}\vec{\na}\mathcal{V}(\vec{u})\cdot\hat{\vec{g}}\mathcal{V}(\vec{u})-i\frac{(L-\bar{z})^2}{2\o}\hat{\vec{g}}\cdot\vec{\na}\mathcal{V}(\vec{u})\notag\\
&-\vec{w}\cdot\hat{\vec{g}}\mathcal{V}(\vec{u})(L-\bar{z})-\frac{(L-\bar{z})^2}{2\o}\vec{k}\cdot\hat{\vec{g}}\mathcal{V}(\vec{u})\Bigg]\,,
\end{align}
where $\vec{u}=\vec{y}-\bar{\vec{x}}$, $\vec{w}=\frac{\vec{y}+\bar{\vec{x}}}{2}$, and $L$ is the matter extension. The effective in-medium scattering potential $\cV$ is often referred to as a dipole potential, and is given by
\begin{align}
\cV(\q) = -\mathcal{C}\rho \left(|v(\q)|^2 -\delta^{(2)}(\q) \int_\k |v(\k)|^2\right) \,,
\end{align}
where $v(\q)=v(\q,0, z)$, and in the case of the GW model $v(\q)=-\frac{g^2}{\q^2+\m^2(z)}$ with $\m(z)$ being constant. Here, we have also introduced an overall color coefficient $\mathcal{C}=\frac{N_c}{2C_{\bar R}}$. In turn, outside the medium ($z_f>\bar{z}>L$), the broadening two-point function is trivial, and can be obtained from \eqref{S2} in the limit $\mathcal{V}\to0$, reducing to $S_2\left(\vec{k},\vec{k},\infty; \vec{y}, \bar{\vec{x}},L\right)=e^{-i\tvec{k}\cdot \vec{u}}$. 

The emission kernel is similar to the broadening two-point function in its structure, but is harder to evaluate for the general potential $\cV$. Following \cite{Barata:2022krd}, one can write the medium average of two Wilson lines up to the leading order in gradient corrections as 
\begin{align}
    &\left< \cP \exp \left\{i \int_z^{\bar{z}} d\t \, T^a \tilde{v}^a(\mathbf{r},\t) \right\} \cP \exp \left\{ - i \int_z^{\bar{z}} d\t \, T^b \tilde{v}^b (\x_{in},\t) \right\} \right> \notag
    \\ & \hspace{2cm} \simeq \exp\left\{ - \int_z^{\bar{z}} d\t \left(1+\frac{\mathbf{r}+\x_{in}}{2}\cdot\hat{\g}\right) \cV(\mathbf{r}-\x_{in}) \right\}  \, ,
\end{align}
and the kernel can be expressed as the corresponding path integral, {\it c.f.} with the homogeneous case \cite{Zakharov:1996fv, Casalderrey-Solana:2007knd, Mehtar-Tani:2012mfa,Blaizot:2012fh}. Thus, it takes the form
\begin{align}
\label{Kdef}
\mathcal{K}\left(\vec{y},\vec{x}_{in},\bar{z}; \vec{x},\vec{x}_{in}, z\right)&= \int_\x^\y D\rn \,  e^{\frac{i\omega}{2} \int d\tau \, \dot{\rn}^2 -\int_z^{\bar z} d\tau \, \left(1+\frac{\rn(\tau)+\x_{in}}{2}\cdot \hat \g \right) \cV\left(\rn(\tau)-\x_{in}\right) }\,,
\end{align}
where the leading gradient corrections are also included. Here we notice, that when both of the longitudinal arguments in the kernel are outside the medium, it also becomes vacuum-like
\begin{align}
\mathcal{K}\left(\vec{y},\vec{x}_{in},\bar{z}; \vec{x},\vec{x}_{in}, z\right)|_{z>L}&= \int_{\k} e^{i\k\cdot(\y-\x)}e^{-i\frac{\k^2}{2\o}(\bar{z}-z)}\,,
\end{align}
while for $\bar{z}>L>z$ it can be expressed as a convolution
\begin{align}
\mathcal{K}\left(\vec{y},\vec{x}_{in},\bar{z}; \vec{x},\vec{x}_{in}, z\right)|_{\bar{z}>L>z}&= \int_{\k} e^{i\k\cdot\y}e^{-i\frac{\k^2}{2\o}(\bar{z}-L)}\mathcal{K}\left(\k,\vec{x}_{in},L; \vec{x},\vec{x}_{in}, z\right)\,,
\end{align}

Thus, it is convenient to split the full distribution into three portions, coming from the three integration regions: $\bar{z}<L$, $z<L<\bar{z}$, and $L<z$. We will refer to these regions as "In-In", "In-Out", and "Out-Out" respectively, see e.g. \cite{Casalderrey-Solana:2007knd}. One readily finds that the in-medium contribution is given by
\begin{align}
\label{Niningeneral}
    &2(2\pi)^3 \omega E \frac{dN^{\rm In-In}}{d\omega dE d^2\k}=\frac{2\alpha_sC_F}{\omega^2} \re \int_0^L d\bar{z}\int_0^{\bar{z}} dz \int_{\x_{in}, \y }  e^{-i\tvec{k} \cdot \y}|J(\x_{in})|^2 \cP_{L-\bar{z}}(\y)\nn 
    &\hspace{0.9cm}\times\Bigg\{ \Bigg[1+i\frac{(L-\bar{z})^2}{2\o}\hat{\vec{g}}\mathcal{V}(\y)\cdot\vec{\na}_{\y}-\frac{\y+2\x_{in}}{2}\cdot\hat{\vec{g}}\mathcal{V}(\y)(L-\bar{z})\Bigg]\vec{\na}_{\y}-\hat{\vec{g}} \mathcal{V}(\y) (L-\bar{z}) \Bigg\}\,\notag\\ 
    &\hspace{1.75cm}\cdot \vec{\na}_\x\mathcal{K}\left(\y + \x_{in},\x_{in},\bar{z}; \vec{x},\x_{in}, z\right)\Bigg|_{\vec{x}=\bar{\vec{x}}=\x_{in}}\,,
\end{align}
where we have substituted \eqref{S2}, and integrated by parts. We have also introduced a broadening probability
\begin{align}
\cP_{L-\bar{z}}(\y)=e^{-\cV(\y)(L-\bar{z})}\left[1-\frac{i(L-\bar{z})^3}{6\o}\vec{\na}\cV(\y)\cdot\hat{\vec{g}}\cV(\y)\right]\,,
\end{align}
which controls the broadening process for a narrow initial distribution \cite{Barata:2022krd}. One may notice that the structure in the second line of \eqref{Niningeneral} resembels the shift operator introduced in \cite{Barata:2022krd}. Indeed, the emission kernel $\cK$ convoluted with the source of the parent parton $J$ serves as a source for the consequent broadening of the emitted gluon. However, the broadening structure is more involved now, since the effective initial distribution is sensitive to the center of mass position of the color dipole $\w$ in \eqref{S2}, and new terms, absent in the results of \cite{Barata:2022krd}, appear. In turn, the In-Out contribution reads
\begin{align}
\label{Ninoutgeneral}
    &2(2\pi)^3 \omega E \frac{dN^{\rm In-Out}}{d\omega dE d^2\k}=\frac{2\alpha_sC_F}{\omega} \re \int_0^{L} dz \int_{\x_{in}} e^{i\k\cdot\x_{in}}|J(\x_{in})|^2\notag\\ 
    &\hspace{2.5cm} 2\frac{\k}{\k^2}\cdot\vec{\na}_\x \mathcal{K}\left(\k,\vec{x}_{in},L; \vec{x},\vec{x}_{in}, z\right)\Bigg|_{\vec{x}=\x_{in}}\,,
\end{align}
where we have explicitly evaluated the $\bar{z}$-integral, noticing that the vacuum kernel is regulated to decay at the infinity. Finally, one may notice that the "Out-Out" contribution corresponds to the physical situation when the emission happens outside the matter, both in direct and conjugated amplitudes. Here, we will omit it, focusing solely on the medium induced part of the spectrum.

\section{Medium induced spectrum in the harmonic approximation}\label{sec:medium_spectrum_harmonic}
In this section, we will focus on the medium induced gluon distribution at the leading order in gradients. However, even in the homogeneous limit, the resummed medium induced gluon distribution has highly non-trivial structure, and cannot be treated analytically. Here, we will rely on additional approximations, which considerably simplify the results and are commonly used in the homogeneous case, see e.g. \cite{Casalderrey-Solana:2007knd, Mehtar-Tani:2012mfa,Blaizot:2012fh,Andres:2020vxs,Barata:2021wuf}. 

First, we ignore the initial state effects, taking a broad source approximation and setting $|J(\x_{in})|^2=f(E)\delta^{(2)}(\x_{in})$. This allows us to simplify $\cK$, and relate the particle distribution $dN$ with the medium induced gluon spectrum $dI$ in the regular way. Thus, we write
\begin{align}
  (2\pi)^2 \omega E \frac{dN}{d\omega dE d^2\k} = (2\pi)^2\omega\frac{dI}{d\omega d^2\k}\, E\frac{dN_0}{ dE}   \,,
\end{align}
where $E\frac{dN_0}{ dE}=\frac{1}{2(2\pi)}f(E)$. The only dependence of the full final state distribution on the initial condition comes from the quark spectrum and can be trivially factorized. This effective form for the particle distribution can be argued to follow from the QCD factorization for soft radiation.

We further notice that the path integral in \eqref{Kdef} cannot be performed analytically in the general case, see e.g. \cite{Casalderrey-Solana:2007knd, Mehtar-Tani:2012mfa,Blaizot:2012fh,Andres:2020vxs,Barata:2021wuf} for discussions, and even for the specified GW potential we are using here as an example. However, to illustrate the leading gradient effects more explicitly, we find it instructive to consider a tractable model. To do so, let us assume that the effective scattering potential is quadratic, $\cV(\y)=\frac{\hat q}{4}\y^2$, focusing on the so-called harmonic approximation. This effective potential can be understood as a separate medium model, but often it is considered as an approximation to other models, such as the GW one, see e.g. \cite{Barata:2021wuf} for a recent discussion. The proportionality coefficient $\hat q$ is commonly referred to as the jet quenching parameter. It is closely related with the broadening process, and, in the general case, it is defined as $\hat{q}=\frac{\pa}{\pa L}\langle \p^2\rangle$. For instance, for the GW model, one finds that $\hat q = \frac{\mu^2 \chi}{L} \log \frac{\L^2}{\mu^2}$ with $\chi \equiv \frac{\mathcal{C}\,g^4\,\rho}{4\pi\mu^2}L$ being the medium opacity. The Coulomb logarithm is divergent and regulated by a momentum scale $\L$, which can be understood as a free parameter of the medium model. Taking $\hat{q}$, as in the GW model, one may consider the harmonic approximation potential for $\cV(\y)$ as the leading contribution at small $\m|\y|$ with a regulated logarithm. Here, we will not discuss this relation or its phenomenological relevance further, utilizing the harmonic potential and explicit form of $\hat{q}$ in the GW model only as illustrations of our general results \eqref{Niningeneral} and \eqref{Ninoutgeneral}.

Now, we can further simplify the emission spectrum, expanding the broadening two-point function and emission kernel as $\cK\simeq \cK^{(0)} + \delta \cK $ and $\cP\simeq \cP^{(0)} +\delta \cP$, where the perturbations are linear in gradients. In the absence of gradients, the path integral in \eqref{Kdef} becomes Gaussian in the harmonic approximation. Thus, it can be readily evaluated \cite{Casalderrey-Solana:2007knd, Mehtar-Tani:2012mfa,Blaizot:2012fh,Barata:2021wuf}, resulting in
\begin{align}
 \cK^{(0)}(\y,\bar z;\x,z)\equiv\cK^{(0)}(\y,0,\bar z;\x,0,z)&\notag\\
 &\hspace{-4cm}\simeq \frac{A_{\bar{z}z}}{\pi i} \exp \left\{iA_{\bar{z}z}B_{\bar{z}z}(\y^2+\x^2)-2i A_{\bar{z}z}\y\cdot \x\right\} \,,
\end{align}
where we have introduced $A_{\bar{z}z}= \frac{\omega \Omega}{2 \sin(\Omega (\bar z-z))}$ and $B_{\bar{z}z}= \cos(\Omega (\bar z-z))$
with $\Omega=\frac{1-i}{2}\sqrt{\frac{\hat q}{\omega}}$, and set $\x_{in}=0$. Treating the leading gradient correction to the potential in \eqref{Kdef} perturbatively, we further find that
\begin{align}
\label{eq:Dyson_K}
&\d\cK \left(\vec{y},\bar{z}; \vec{x}, z\right)\equiv\cK \left(\vec{y},\bar{z}; \vec{x}, z\right) - \cK^{(0)}\left(\vec{y},\bar{z}; \vec{x}, z\right)\notag\\
&\hspace{4cm}= -\frac{1}{2} \, \vec{g}\cdot\int_{\w}\int_z^{\bar z} ds \,    \w\,\cK^{(0)}(\vec{y},\bar{z}; \vec{w}, s)\cV(\w, s)\cK^{(0)}(\vec{w},s; \vec{x}, z)\, ,
\end{align}
where $\hat{\g}$, acting solely on $\hat q$, has been replaced with $\g=\frac{1}{\hat{q}}\hat{\g}\hat{q}$. Similarly, in the harmonic approximation, the broadening probability is also Gaussian, and reads
\begin{align}
   \cP^{(0)}_z(\p)&\equiv \int_{\x} e^{-i\p\cdot\x}e^{-\mathcal{V}\left(\x\right)z} = \frac{4\pi}{\hat q z} e^{-\frac{\p^2}{\hat q z}}\, ,
\end{align}
while the leading gradient correction is given by
\begin{align}
   \delta \cP_z(\p)&\equiv\int_{\x} e^{-i\p\cdot\x}e^{-\mathcal{V}\left(\x\right)z}\left[-\frac{iz^3}{6\omega } \na\mathcal{V}\left(\x\right)\cdot\g \mathcal{V}\left(\x\right)\right]=\frac{4\pi}{\hat q z} e^{-\frac{\p^2}{\hat q z}} \, \frac{z}{6\omega } \left(\frac{\p^2-2\hat{q} z}{\hat q z}\right) \g\cdot \p \,.
\end{align}

Expanding the medium induced soft gluon spectrum to the leading order in $\g$ we can write it with these notations as
\begin{align}
\label{Isum}
&\omega\frac{dI}{d\omega d^2\k}= \omega \frac{dI_0}{d\omega d^2\k}+ \omega\frac{dI_\cP}{d\omega d^2\k}+ \omega\frac{dI_\cK}{d\omega d^2\k}+ \omega\frac{dI_{\hat{\mathsf S }}}{d\omega d^2\k}\,,
\end{align}
where the gradient corrections are grouped by their origin for convenience. While the notations for $dI_\cP$ and $dI_\cK$ are clear, they correspond to the gradient terms coming from $\d\cP$ and $\d\cK$, the last contribution in \eqref{Isum} has a subscript $\hat{\mathsf S}$. This object corresponds to the terms in the second line of \eqref{Niningeneral}, and it can be thought of as a generalization of the shift operator, adjusting the argument of the initial distribution by a gradient contribution in the case of simple momentum broadening in \cite{Barata:2022krd}. Here, the effective initial distribution of the emitted gluon is given by a convolution of $\cK$ with $J$, while some additional terms appear due to the dependence of the answer on the center of mass position of the color dipole. Finally, we notice that, since the presence of the gradients is naturally attached to the presence of the medium, only $dI_\cK$ may be non-trivial in the In-Out region.

\section{The gradient corrections}
\label{sec:gradients}
Now, we are in the position to evaluate the novel contributions. As a warm-up exercise, we start by considering the standard spectrum $dI_0$, appearing in the case of homogeneous medium. Some of the intermediate calculations, which are not shown explicitly here, can be found in great detail in the literature, see e.g. \cite{Barata:2021wuf}. 

As has already been mentioned, the two different integration regions result in two contributions to the gluon spectrum. From \eqref{Niningeneral} one readily finds that in the absence of the gradients the In-In part is given by 
\begin{align}\label{eq:Inin_bdmps}
(2\pi)^2 \omega\frac{dI^{\rm In-In}_0}{d\omega d^2\k}&=\frac{2\alpha_sC_F}{\omega^2}\re \int_0^L d\bar{z}\int_0^{\bar{z}} dz \int_{ \y}  e^{-i\tvec{k}\cdot \y} \cP^{(0)}_{L-\bar{z}}(\y) \vec{\na}_\y \cdot \vec{\na}_\x\cK^{(0)}\left(\vec{y},\bar{z}; \vec{x}, z\right)\Bigg|_{\vec{x}=0} \nn
&= \frac{4\alpha_sC_F}{\omega } \re \int_{0}^L d\bar z  \, \frac{e^{\frac{\k^2 \cT_{\bar z}}{i-Q_{\bar{z}}^2\cT_{\bar{z}}}}}{Q_{\bar{z}}^2\cT_{\bar{z}} -i} \, ,
\end{align}
where we have introduced shorthand notations
$Q_{\bar{z}}^2 = \hat{q} (L-\bar{z})$ and $\cT_z = \frac{\tan(\Omega z)}{2\omega \Omega}$. This is the simplest analytical form, which can be obtained without switching to the numerical tools, which we will do later.

In turn, the In-Out part follows from \eqref{Ninoutgeneral}, and in the absence of the gradients we readily find the standard answer
\begin{align}
(2\pi)^2 \omega\frac{dI^{\rm In-Out}_0}{d\omega d^2\k}&= \frac{2\alpha_sC_F}{\omega}\re \,  \int_0^L d  z \,  \int_{ \y} e^{-i \k \cdot \y} \,  2\frac{\k}{\k^2}   \cdot \vec{\na}_\x \cK^{(0)}\left(\vec{y},L; \vec{x}, z\right)\Bigg|_{\vec{x}=0} \nn
&=-\frac{8\alpha_sC_F}{\k^2} \, \re \left(1-e^{-i \k^2 \cT_L}\right)\, .
\label{InOut0}
\end{align}

The $dI_\cP$ term accounts for the gradient effects on the late time evolution of the gluon. As a consequence, it only has support in the In-In region. Explicitly, it can be written as 
\begin{align}
\label{eq:Icp_1}
(2\pi)^2\omega \frac{dI^{\rm In-In}_\cP}{d\omega d^2\k}&=\frac{2\alpha_sC_F}{\omega^2}\re \int_0^L d\bar{z}\int_0^{\bar{z}} dz \int_{ \y}  e^{-i\tvec{k}\cdot \y} \delta \cP_{L-\bar{z}}(\y) \vec{\na}_\y \cdot \vec{\na}_\x\cK^{(0)}\left(\vec{y},\bar{z}; \vec{x}, z\right)\Bigg|_{\vec{x}=0} \nn 
&= \frac{8\alpha_sC_F \pi }{3\hat q\omega^2}\re \int_0^L d\bar{z}\int_\q \, i\,  e^{-\frac{\q^2}{Q_{\bar{z}}^2}}   \left(\frac{\q^2-2Q_{\bar{z}}^2}{Q_{\bar{z}}^2}\right) \g\cdot \q \,   e^{-i (\k-\q)^2\cT_{\bar z} } \, ,
\end{align}
where the remaining integral over $\q$ is Gaussian, and we find
\begin{align}
\label{Pinin}
(2\pi)^2\omega \frac{dI^{\rm In-In}_\cP}{d\omega d^2\k}&= \frac{2\alpha_sC_F}{3\hat q\omega^2} (\g\cdot \k) \, \re  \int_0^L d\bar{z} \, e^{\frac{\k^2 \cT_{\bar{z}}}{i-Q_{\bar{z}}^2\cT_{\bar{z}}}} Q_{\bar{z}}^6 \cT^2_{\bar{z}} \; \frac{2i+(\k^2-2Q_{\bar{z}}^2)\cT_{\bar{z}}}{(i-Q_{\bar{z}}^2\cT_{\bar{z}})^4} \, .
\end{align}

The $I_\cK$ part of the spectrum is the only gradient correction, having support in both the In-In and In-Out regions, and it results in two contributions.  The In-In term is given by
\begin{align}
&(2\pi)^2\omega \frac{dI^{\rm In-In}_\cK}{d\omega d^2\k}=  \frac{2\alpha_s C_F}{\o^2}\re \int_0^L d\bar{z}\int_0^{\bar{z}} dz \int_{ \y}  e^{-i\tvec{k}\cdot \y} \cP^{(0)}_{L-\bar{z}}(\y) \vec{\na}_\y\cdot \vec{\na}_\x \delta \cK\left(\vec{y},\bar{z}; \vec{x}, z\right)\Bigg|_{\vec{x}=0} \, ,
\end{align}
and, after some algebra, it can be written as
\begin{align}
&(2\pi)^2\omega \frac{dI^{\rm In-In}_\cK}{d\omega d^2\k}=-\frac{\alpha_s C_F\, \hat{q}}{4\pi\o}\, \re\int_0^L d\bar z \int_0^{\bar z} dz \int_{\q,\w}   \frac{e^{-i  \cT_{\bar{z}-z} \q^2}}{B_{\bar{z}z}} \, \cP^{(0)}_{L-\bar{z}}(\k-\q)  \nn
&\times e^{\frac{i}{4} \frac{\w^2}{\cT_z}} \,  (\g \cdot \w) \,  (\q\cdot \w)\,\exp \left\{-i\left(\frac{\q \cdot \w}{B_{\bar{z}z}}-\frac{A_{\bar{z}z}(B_{\bar{z}z}^2-1)}{B_{\bar{z}z}}\w^2\right)\right\}\,,
\end{align}
where we have changed the order of integration, and renamed the intermediate position $s$ by $z$, while the initial $z$-integration has been performed. The transverse integrals can be evaluated analytically, resulting in
\begin{align}
&(2\pi)^2\omega \frac{dI^{\rm In-In}_\cK}{d\omega d^2\k}=\frac{32\alpha_s C_F\, \hat{q}}{\o} \left(\g\cdot \k \right)\re  \int_0^L d\bar z \int_0^{\bar z} dz \,e^{\frac{\k^2 \cT_{\bar{z}}}{i-Q_{\bar{z}}^2\cT_{\bar{z}}}}\cT_z^2\notag\\
    & \hspace{1cm} \times\frac{3Q_{\bar{z}}^4 \cT_z^2- B_{\bar{z}z}^2 C_{\bar{z}z}^2 \left(i-Q_{\bar{z}}^2 \cT_{\bar{z}-z}\right)^2 - 2i B_{\bar{z}z} C_{\bar{z}z} \cT_z \left(\vec{k}^2+ Q_{\bar{z}}^2 + i Q_{\bar{z}}^4 \cT_{\bar{z}-z}\right)}{16C^4_{\bar{z}z}\left(i-Q^2_{\bar{z}}\cT_{\bar z}\right)^4B_{\bar{z}z}} \, ,
\end{align}
where $C_{ts}= \frac{\cos\left(\O t\right)}{\cos\left(\O s\right)}$. Finally, one may notice that in this form the $z$-integration is sufficiently simple, resulting in
\begin{align}
\label{Kinin}
&(2\pi)^2\omega \frac{dI^{\rm In-In}_\cK}{d\omega d^2\k}=\frac{8\alpha_s C_F}{3\hat{q}\o^2} \left(\g\cdot \k \right)\re  \int_0^L d\bar z \,\frac{e^{\frac{\k^2 \cT_{\bar{z}}}{i-Q_{\bar{z}}^2\cT_{\bar{z}}}}}{\left(i-Q^2_{\bar{z}}\cT_{\bar z}\right)^4} \frac{\sin^2\left(\frac{\O\bar{z}}{2}\right)}{\cos^2\left(\O\bar{z}\right)}\notag \\
&\hspace{1cm} \times\Bigg\{-\o\O \sin\left(\O\bar{z}\right)+ i \left[\vec{k}^2+i Q_{\bar{z}}^4 \cT_{\bar{z}} + \frac{2\left(\vec{k}^2+Q_{\bar{z}}^2\right)+3iQ_{\bar{z}}^4  \cT_{\bar{z}}}{\cos\left(\O \bar{z}\right)}\right]\sin^2\left(\frac{\O\bar{z}}{2}\right)\Bigg\} \,.
\end{align}

In turn, the In-Out term can be written in the form similar to \eqref{Ninoutgeneral}, reading
\begin{align}
(2\pi)^2\omega \frac{dI^{\rm In-Out}_\cK}{d\omega d^2\k}&= \frac{2 \alpha_s C_F}{\o}\re \,   \int_0^L d  z \,  \int_{ \y} e^{-i \k \cdot \y} \, 2\frac{\k}{\k^2} \cdot \vec{\na}_\x \delta \cK\left(\vec{y},L; \vec{x}, z\right)\Bigg|_{\vec{x}=0}  \, ,
\end{align}
and, after the transverse integrals are performed, it reduces to
\begin{align}
(2\pi)^2\omega \frac{dI^{\rm In-Out}_\cK}{d\omega d^2\k}&=-\frac{4\alpha_s\hat{q}C_F}{ \k^2}\,(\g \cdot \k)\, \re \int_0^L d  z \,  \frac{   \cT^2_z}{C^3_{Lz}} e^{-i \k^2 \cT_L} (iB_{Lz}C_{Lz}+2\k^2 \cT_z)\,.
\end{align}

Finally, we notice that the last $z$-integral can be treated explicitly, leading to
\begin{align}
\label{Kinout}
&(2\pi)^2\omega \frac{dI^{\rm In-Out}_\cK}{d\omega d^2\k}=- \frac{4}{3 \omega^2 }\alpha_s C_F \hat q  \frac{\g \cdot \k}{ \k^2}\, \re    \frac{  e^{-i \k^2 \cT_L} }{\Omega^3}\notag\\
&\hspace{1cm}\times\Bigg[ i \frac{\sin^2\left(\frac{\Omega L}{2}\right) \sin(\Omega L)}{\cos^2(\Omega L)} + \frac{\k^2}{\omega \Omega} (2+ \cos(\Omega L)) \frac{\sin^4\left(\frac{\Omega L}{2}\right)}{\cos^3(\Omega L)} \Bigg]\,.
\end{align}

Now, we can turn to the $dI_{\hat{\mathsf S}}$ contribution. This term has support only inside the medium, and reads
\begin{align}
\label{eq:IS}
(2\pi)^2\omega \frac{dI^{\rm In-In}_{\hat{\mathsf S}}}{d\omega d^2\k}&= \frac{2\alpha_sC_F}{\o^2}\re \int_0^L d\bar{z}\int_0^{\bar{z}} dz \int_{ \y}  e^{-i\tvec{k}\cdot \y} \cP^{(0)}_{L-\bar{z}}(\y)\nn 
     &\times\Bigg\{ \Bigg[i\frac{(L-\bar{z})^2}{2\o}\hat{\g}\mathcal{V}(\y)\cdot\vec{\na}_{\y}-\frac{\y}{2}\cdot\hat{\g}\mathcal{V}(\y)(L-\bar{z})\Bigg]\vec{\na}_{\y}-\hat{\g} \mathcal{V}(\y) (L-\bar{z}) \Bigg\}\,\notag\\ 
     &\cdot \vec{\na}_\x\cK^{(0)}\left(\vec{y},\bar{z}; \vec{x}, z\right)\Bigg|_{\vec{x}=0} \, .
\end{align}
Evaluating the $z$-integral, we bring it to a particularly simple form
\begin{align}
(2\pi)^2\omega \frac{dI^{\rm In-In}_{\hat{\mathsf S}}}{d\omega d^2\k}&= \frac{\alpha_s C_F\,\hat{q}}{2\pi\o}\re \int_0^L d\bar{z}  \int_{\y}  e^{-i\tvec{k}\cdot \y}e^{\frac{i }{4\cT_{\bar{z}}}\y^2} \cP^{(0)}_{L-\bar{z}}(\y) \nn 
     &\times(L-\bar z)\Bigg[i-\frac{\y^2}{4\cT_{\bar{z}}} \left(1+ \frac{L-\bar{z}}{2\omega \cT_{\bar{z}}}\right)\Bigg] (\g \cdot \y) \, .
\end{align}

The remaining Fourier transformation can be performed analytically, and one finds
\begin{align}
\label{Sinin}
(2\pi)^2\omega \frac{dI^{\rm In-In}_{\hat{\mathsf S}}}{d\omega d^2\k}&= -\frac{4\alpha_s C_F}{\o}(\g\cdot \k)\re \, \int_0^L d\bar{z} \,  Q_{\bar{z}}^2 \, \cT_{\bar{z}}^2\,e^{\frac{\k^2\cT_{\bar{z}} }{i-Q_{\bar{z}}^2 \cT_{\bar{z}}}}\nn 
&\times \frac{\left[ i \left(1+ \frac{L-\bar{z}}{2\omega \cT_{\bar{z}}}\right) \left(\k^2 \cT_{\bar{z}}-2 Q_{\bar{z}}^2 \cT_{\bar{z}}+2 i\right)-\left(i-Q_{\bar{z}}^2
   \cT_{\bar{z}}\right)^2\right]}{\left(i-Q_{\bar{z}}^2 \cT_{\bar{z}}\right)^4}\,.
\end{align}
Here, one should also notice that all the final expressions \eqref{Pinin}, \eqref{Kinin}, \eqref{Kinout}, and \eqref{Sinin} are proportional to $(\g\cdot\k)$, leading to the term $(\g\cdot\k)\,dI_1$ in \eqref{spectrumintro}. Thus, on average, the emitted gluons have their momenta aligned with the direction of $\tvec{g}$. 

Turning to the further analysis of the analytic results above, we notice that the gradient corrections to the gluon spectrum have some similarity with \eqref{eq:Inin_bdmps} and \eqref{InOut0}. Indeed, the same exponential factors and similar polynomials of $Q_{\bar z}^2$ and $\cT_{\bar z}$ appear in the results of this section, leaving the possibility that the gradient corrections can be obtained with an operator acting on the leading contributions. Another immediate observation is that the leading corrections depend on the same dimensionful scales, while the new scale introduced by the gradients is factorized. We leave these opportunities to deeper understand the qualitative properties of the results for future work, and focus on the overall behavior of the full spectrum in the next section.

\section{The spectrum and its properties}\label{sec:results}

In this section, we will focus on the properties of the medium induced soft gluon spectrum and its gradient corrections in the harmonic approximation. Despite all the simplifications, the last $\bar{z}$-integration in the In-In contributions, \eqref{eq:Inin_bdmps} and \eqref{InOut0}, is pretty involved even in the homogeneous limit, and usually treated numerically, see e.g. \cite{Casalderrey-Solana:2007knd, Mehtar-Tani:2012mfa, Blaizot:2012fh,Andres:2020vxs,Barata:2021wuf,Feal:2018sml,Feal:2019xfl,Caron-Huot:2010qjx}. Here, we choose a set of phenomenologically relevant parameters \cite{Barata:2022krd}, although the considered limit of the medium induced gluon spectrum is oversimplified and used for illustrative purposes. First, we define the medium to have $L=5$ fm and $T=0.3$ GeV, and assume that $\tvec{\na}T\lesssim T^2$ for hydrodynamically evolving matter. If we treat the harmonic oscillator approximation as our model, then the only other parameter is $\hat q$, and its characteristic value can be chosen to be about $\hat{q}\simeq 1\,\text{GeV}^2\cdot\text{fm}^{-1}$, see e.g. \cite{Mehtar-Tani:2012mfa,Blaizot:2012fh,Barata:2021wuf}. On the other hand, to illustrate the logarithmic dependence in $\hat q$, which affects the definition of the gradient vector $\g$, one may attempt at treating the harmonic effective potential as a crude approximation of the GW model. For this purpose, we set $\chi=2.75$, $\mu=0.6$ GeV, and choose $\Lambda^2=E \mu$ with characteristic jet energy of 100 GeV, {c.f.} with \cite{Sadofyev:2021ohn, Barata:2022krd}. Then, one readily finds that $\hat{q}=\frac{\chi \m^2}{L}\log \frac{E}{\m}\simeq 1\, \text{GeV}^2\cdot\text{fm}^{-1}$, and the two values coincide. Finally, turning to the gradient vector, we notice that the form of $\g$ is controlled by the powers of temperature, entering into the scaling $\hat{q}\sim T^3\log\frac{\Lambda^2}{\m^2}$ \cite{Barata:2022krd}. For instance, for the naive scaling of the jet quenching parameter $\g = 3\frac{\tvec{\na} T}{T}$, while if one takes into account the logarithmic dependence as in the GW model with the given form of the cutoff it reads $\g = \frac{\tvec{\na} T}{T} \left(3-\log^{-1} \frac{E}{\mu}\right)$. However, the characteristic value of $\Lambda$ corresponding to our choice of $\hat q$ is such that the logarithmic factor in the GW model is sufficiently large, resulting in $\g\simeq2.8 \frac{\tvec{\na} T}{T}$, and, for simplicity, we will use $\g = 3\frac{\tvec{\na} T}{T}$ for all our estimates.

The medium induced soft gluon spectrum has an angular dependence controlled by $\g\cdot \k$, and we will focus on the two limiting cases, when the angle between the two vectors, $\theta$, is either 0 or $\pi$. We will measure the gluon frequencies with respect to the critical medium frequency $\omega_c \equiv \hat q L^2\simeq 125$ GeV, which in the case of no gradients can be identified with the typical frequency for gluons with formation length of the order of $L$. We will also introduce a dimensionless gradient parameter $\gamma_T=|\vec{\na}T/T^2|$, which controls the strength of the hydrodynamic gradients and distribution anisotropy.

\begin{figure}[h!]
    \centering
    \includegraphics[width=\textwidth]{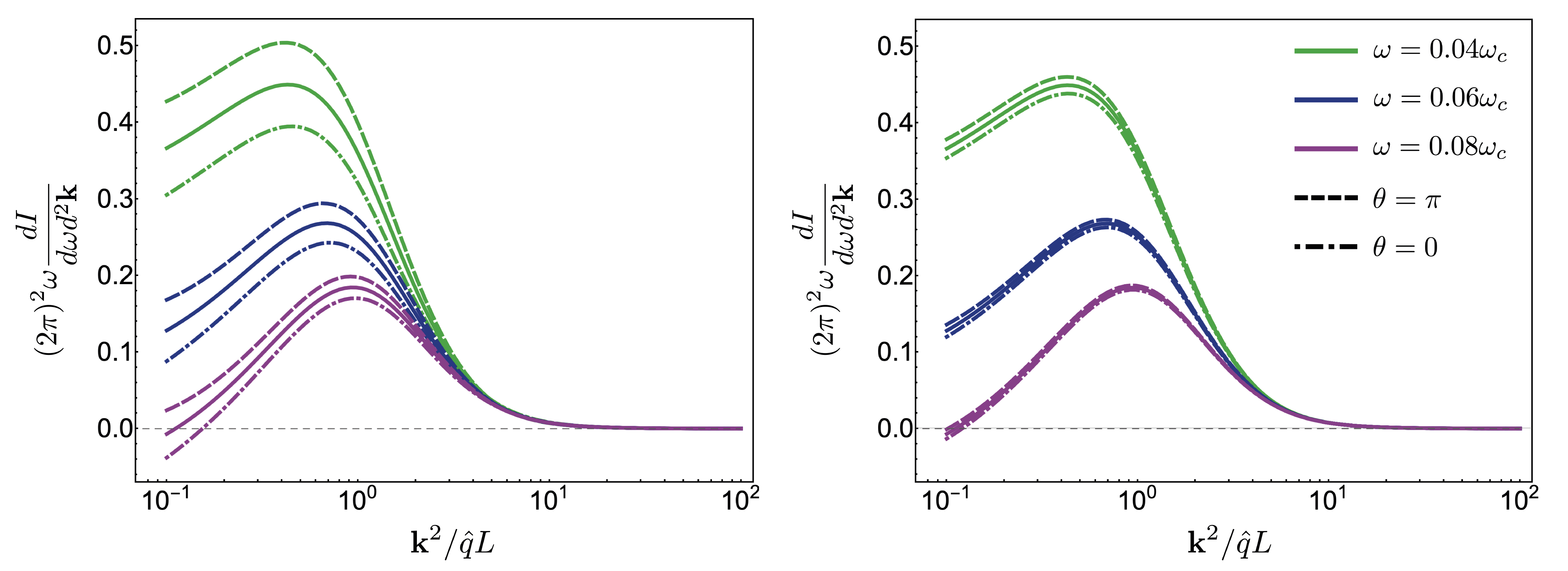}
    \caption{ The medium induced soft gluon spectrum is given for three gluon energies, $\omega=0.04\, \omega_c$, $\omega=0.06 \, \omega_c$, and $\omega=0.08 \, \omega_c$. The solid lines denote the spectrum in the homogeneous limit. The dashed and dash-dotted lines correspond to the full spectrum with gradients along ($\theta=0$) and opposite to ($\theta=\pi$) the direction of $\k$ respectively. The gradients are quantified with $\gamma_T=0.05$ (left) and $\gamma_T=0.01$ (right).}
    \label{fig:fig2}
\end{figure}

In Fig.~\ref{fig:fig2}, we show the full spectrum up to first order in gradient corrections for $\omega=  \, 0.04\,\omega_c$, $\omega=  \, 0.06\,\omega_c$, and $\omega=0.08 \, \omega_c$, further differentiating for $\gamma_T= 0.05$ (left) and $\gamma_T=0.01$ (right). For $\theta=0$, the gradient effects suppress the gluon radiation at small values of $\k$, while when $\theta=\pi$, it is enhanced. One can notice that the gradient effects in Fig.~\ref{fig:fig2} become stronger for softer gluons, and may be substantial even for sufficiently small $\gamma_T$. This behavior is in line with the properties of the gradient effects in broadening \cite{Barata:2022krd}, where the anisotropic contributions are suppressed by the energy of the leading parton. Since the energy of soft emitted gluons is smaller than the energy of the leading parton, the gradient effects become more important. However, one should notice, that very soft gluons lose their energy on shorter timescales, and the single gluon spectrum cannot describe the evolution of the system reliably in this case. 

\begin{figure}[h!]
    \centering
    \includegraphics[width=\textwidth]{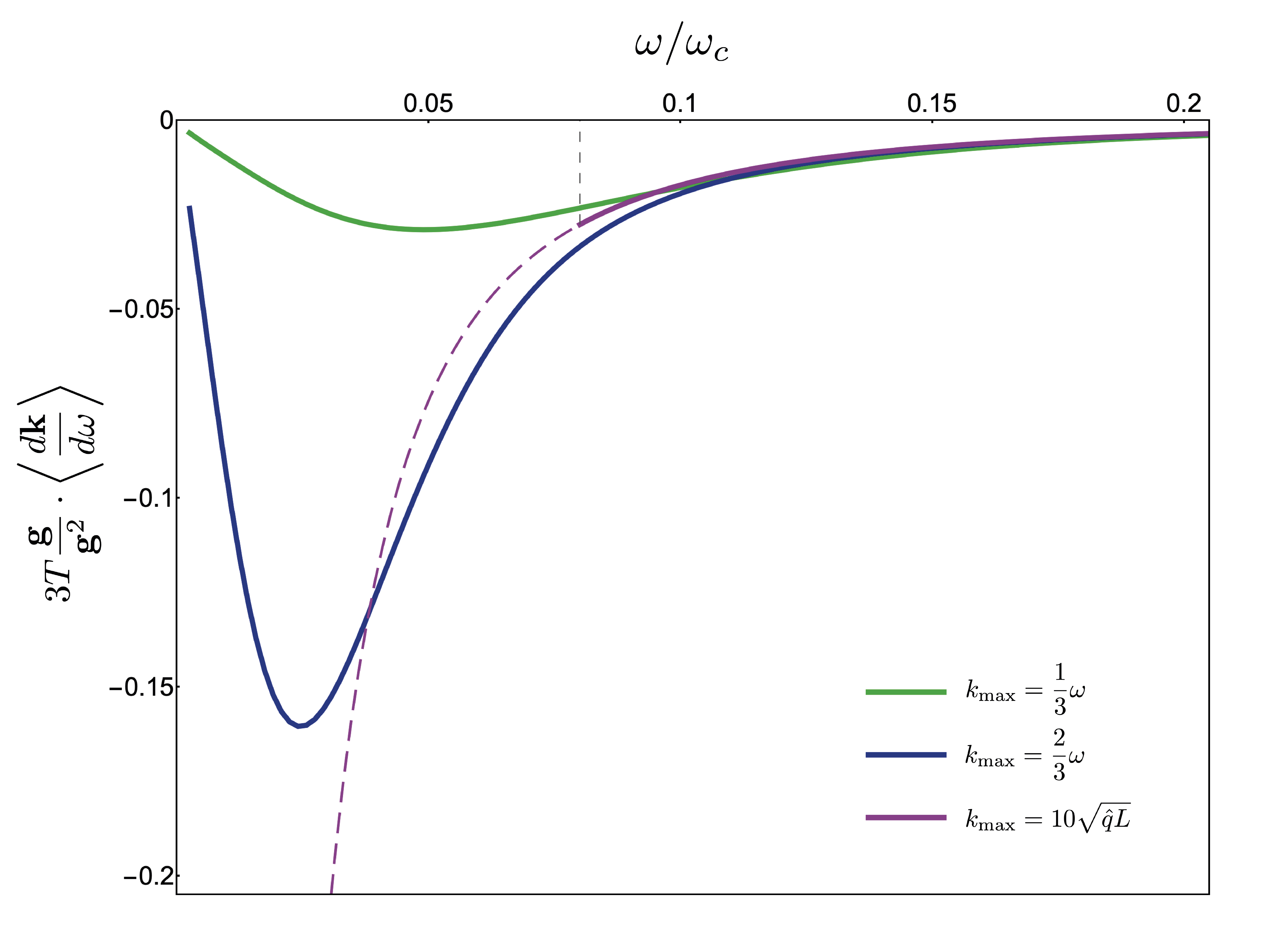}
    \caption{The differential average transverse momentum is given by its dimensionless projection $3T\frac{\g}{\g^2}\cdot\left\langle\frac{d\k}{d\o}\right\rangle$ for three choices of the upper cutoff $k_\max$ in \eqref{kaverage}. The curve corresponding to the $\o$-independent choice of $k_\max$ is shown with a solid line for sufficiently large energies $\o>\sqrt{20\hat{q}L}$, and continued with a lighter dashed line beyond that point. The particular limiting value (indicated with a vertical gray dashed line) corresponds to the point along the spectrum plot, where it is visually close to zero, $\k^2=20\hat{q}L$. The remaining parameters are taken to be the same as in Fig.~\ref{fig:fig2}.}
    \label{fig:kt}
\end{figure}

In quantifying the effect of the resulting anisotropy in the medium induced radiation, one may also consider integral characteristics of the spectrum, such as its moments, see e.g. \cite{Antiporda:2021hpk,Sadofyev:2022hhw,Andres:2022ndd,Barata:2022krd}. Indeed, the gradient corrections to the spectrum are proportional to $\g\cdot \k$, being otherwise functions of $\k^2$, and, thus, its directional odd moments are non-zero, quantifying the average transverse momentum transmitted with the gluons. To illustrate this point, we focus on the differential average transverse momentum, defined as
\begin{align}
\left\langle\frac{d\k}{d\o}\right\rangle \equiv  \int_\G\, d^2\k \, \k \,\frac{dI}{d\omega d^2\k} = \frac{1}{2}\g\int_\G \, d^2\k \, \k^2 \,\frac{dI_1}{d\omega d^2\k}\,,
\label{kaverage}
\end{align}
where $\G$ is a particular phase space region, which should be specified to interpret the physical meaning of $\left\langle\frac{d\k}{d\o}\right\rangle$. Here, we set the lower limit of $|\k|$-integration to zero, pushing the applicability of the harmonic approximation to the limit. However, one may readily check that the numerical results are only weakly affected if we would require $|\k|>\m$. We further focus on the three particular upper cutoffs: $k_\max=10\sqrt{\hat{q}L}$, $k_\max=\frac{2}{3}\o$, and $k_\max=\frac{1}{3}\o$, plotting the numerical results for $3T\frac{\g}{\g^2}\cdot\left\langle\frac{d\k}{d\o}\right\rangle$ in Fig.~\ref{fig:kt}. Notice that $3T\frac{\g}{\g^2}\cdot\left\langle\frac{d\k}{d\o}\right\rangle$ is independent of the value of $|\tvec{\na}T|$, and normalized to give $\left|\left\langle\frac{d\k}{d\o}\right\rangle\right|$, when its absolute value is multiplied by $\gamma_T$. 

The first choice of $k_\max$ is independent of $\o$, and accounts for all the emitted gluons with $|\k|<10\sqrt{\hat{q}L}$. The soft gluon spectrum quickly goes to zero for large momenta, see Fig.~\ref{fig:fig2}, and a sufficiently large upper cutoff, such as $k_\max=10\sqrt{\hat{q}L}$ can be freely replaced with infinity. However, for smaller gluon energies, the gluons contributing to $\left\langle\frac{d\k}{d\o}\right\rangle$ are not necessarily on-shell (and not eikonal), and \eqref{kaverage}  cannot be used as a measure of the averaged emitted transverse momentum. The two other choices account only for the on-shell  emitted gluons, although slightly pushing the results obtained under the eikonal approximation to the limit. These cutoffs applied in \eqref{kaverage} account only for the gluons emitted within smaller conical segments around the leading parton momentum, simulating jet cones. Moving from harder to softer gluons, we notice that the averaged transverse momentum first grows in absolute value, since the leading gradient effects are suppressed by the gluon energy. Thus, as expected, the softer gluons are more sensitive to the matter anisotropies. However, later the spectrum is slightly depleted, while the maximal transverse momentum of contributing gluons is smaller, and the averaged emitted transverse momentum decreases. This results in the peaks on the curves with $\o$-dependent cutoffs in Fig.~\ref{fig:kt}.

Thus, combining insights from this work and the previous studies for single parton broadening \cite{Sadofyev:2021ohn, Barata:2022krd}, one may summarize the picture of jet evolution in inhomogeneous matter in the following way. The partons inside a jet propagating through the matter broaden and radiate in an anisotropic way. The harder partons in the core of the jet are less affected by these directional effects caused by variations in the medium parameters, while the softer radiation can be affected substantially and possesses an imprint of the medium structure. The distribution of the emitted gluons is highly non-trivial, and, for instance, the differential averaged transverse momentum may be parallel or antiparallel to the temperature gradient, depending on the particular integration limits in \eqref{kaverage}. However, for wide enough conical segments $\left\langle\frac{d\k}{d\o}\right\rangle$ is negative as is illustrated in Fig.~\ref{fig:kt}.

\section{Conclusion and Outlook}\label{sec:conclusion}

In this paper, we have derived the double differential medium induced soft gluon spectrum in a dense static transversely inhomogeneous medium with finite longitudinal extension. The spectrum is obtained within a gradient expansion up to the leading first order. As in the case of the jet momentum broadening~\cite{Barata:2022krd}, the gradient effects only enter the final distributions upon averaging over the stochastic background field configurations. In addition to the general form of the spectrum in \eqref{Niningeneral} and \eqref{Ninoutgeneral}, we have considered its behavior for quadratic $\cV(\y)$, the so-called harmonic approximation, evaluating the path integrals explicitly. In this regime, the full spectrum can be written in a form suitable for numerical simulations, and its structure results in no additional computational complications, comparing to the homogeneous case. This indicates that the full soft gluon spectrum can be implemented in the jet quenching phenomenology and simulations for more realistic model potentials, see e.g. \cite{Barata:2021wuf,Feal:2019xfl,Feal:2018sml} for related discussions in the homogeneous limit.

In the harmonic approximation limit, we numerically evaluate the soft gluon spectrum and its leading odd moment $\left\langle\frac{d\k}{d\o}\right\rangle$, and present our results in Fig.~\ref{fig:fig2} and Fig.~\ref{fig:kt}. The medium gradients distort the softer part of the jet substructure, while the harder radiation is less sensitive to the underlying medium structure. The spectrum is depleted for the transverse gradient vector $\g$ parallel with $\k$, and enhanced for the opposite situation, for most of the phase space. However, there are parametric regions, where the ordering is modified, and the spectrum is enhanced for $\k$ parallel with $\g$. Thus, we find that the medium induced soft gluons are preferably emitted along the temperature gradient but the particular direction depends on the region of the gluon phase space. We further study how the average transverse momentum of the medium induced gluons is distributed with energy and what its fraction stays within a conical segment aligned with the initial leading parton momentum, see Fig.~\ref{fig:kt}.

The present theoretical results should be further supplemented with a well-thought set of jet observables sensitive to the medium structure. However, searching for such observables is a non-trivial task, since they are sensitive to the medium anisotropy and are expected to be contaminated by the soft part of the particle spectrum. To overcome this issue, one may focus on substructure observables with less sensitivity to the correlated soft particles.
Presently, the jet substructure techniques applied in the jet quenching phenomenology are still under development, but it is already possible to gauge the sensitivity of particular observables using simplified models for the jet in-medium evolution. Another potential option is to compute some global jet observable (e.g. jet shape) as a function of rapidity. Since in a real event the gradient effects substantially change with rapidity, the difference between such measurements could provide a better access to the medium structure. We leave the study of these questions for the future work.

One should notice that this work makes only the first steps along the discussion of jet-medium interactions in evolving matter, and our results could be extended in multiple ways. For instance, it would be natural to consider higher order gradient corrections within the framework of the present paper. As have been shown in \cite{Barata:2022utc, Barata:2022wim}, in the case of momentum broadening, such terms may arise at the leading eikonal accuracy, while the directional gradient corrections to the final distribution are suppressed at larger energies. Thus, we expect that such higher order terms may significantly alter the medium induced soft gluon spectrum. Therefore, it would be interesting to derive the medium induced spectrum up to the second order in gradients, although such a calculation will be challenging. 

In exploring the particle spectrum, it is necessary to account for more realistic spacetime profiles of the medium. For example, there is an ongoing effort to implement the flow effects in the pQCD description of jet-matter interactions \cite{Sadofyev:2021ohn,Antiporda:2021hpk,Andres:2022ndd}. However, in the case of the medium induced branching, such studies have so far only been applied to the dilute limit. Including higher opacity corrections and gradient effects in a single jet quenching framework is critical to describe jets in realistically evolving matter, and this is an essential step to further develop the tomographic toolkit in HIC. Such a framework can be implemented along with realistic hydrodynamic and/or kinetic theory simulations of the nuclear matter in HIC, providing further insight into the details of jet-matter interactions, for recent efforts in this direction see \cite{Antiporda:2021hpk,Barreto:2022ulg, Zigic:2021rku}.

\section*{Acknowledgements}
The authors would like to thank C. Andres, N. Armesto, F. Dominguez, Y.-J. Lee, M. Sievert, K. Tywoniuk, and B. Wu for fruitful discussions and comments on the presented results. This work is supported by European Research Council project ERC-2018-ADG-835105 YoctoLHC; by Maria de Maetzu excellence program under project CEX2020-001035-M; by Spanish Research State Agency under project PID2020-119632GB-I00; and by Xunta de Galicia (Centro singular de investigación de Galicia accreditation 2019-2022), by European Union ERDF. The work of A.S. is also supported by the Marie Sklodowska-Curie Individual Fellowship under JetT project (project reference 101032858). JB is supported by the U.S. Department of Energy, Office of Science, Office of Nuclear Physics, under contract No. DE-SC0012704. X.M.L. contribution to this work is supported under scholarship No. PRE2021-097748, funded by MCIN/AEI/10.13039/501100011033 and FSE+.

\bibliographystyle{bibstyle}
\bibliography{references}

\end{document}